\documentclass[a4paper,12pt,draftcls,onecolumn]{IEEEtran}

\usepackage{amssymb}
\usepackage{algorithmic}
\usepackage{algorithm}
\usepackage{verbatim}
\usepackage[mathcal]{euscript}
\usepackage{mathrsfs}
\usepackage{cite}

\ifCLASSINFOpdf
\else

   \usepackage[dvips]{graphicx}
   \graphicspath{{figures/}}
   \DeclareGraphicsExtensions{.eps}
\fi
\usepackage{amsmath}\allowdisplaybreaks
\usepackage{amsmath}
\usepackage[font=footnotesize]{subfig}\allowdisplaybreaks[0]

\newtheorem{theorem}{Theorem}
\newtheorem{proposition}{Proposition}
\IEEEoverridecommandlockouts
\begin{document}
\title{ \bf Performance of the Generalized Quantize-and-Forward Scheme over the Multiple-Access Relay Channel}

\author{Ming Lei and Mohammad Reza Soleymani\\
Department of Electrical and Computer Engineering, Concordia University\\ Montreal, Quebec, Canada\\
Email:m\_lei,msoleyma@ece.concordia.ca}

\setlength{\pdfpageheight}{\paperheight}
\setlength{\pdfpagewidth}{\paperwidth}
\maketitle
\thispagestyle{empty}
\pagestyle{empty}

\maketitle

\begin{abstract}
This work focuses on the half-duplex (HD) relaying based on the generalized quantize-and-forward (GQF) scheme in the slow fading Multiple Access Relay Channel (MARC). We consider the case that the relay has no channel state information (CSI) of the relay-to-destination link. Relay listens to the channel in the first slot of the transmission block and cooperatively transmits to the destination in the second slot. In order to investigate the performance of the GQF, the following steps have been taken: 1)The GQF scheme is applied to establish the achievable rate regions of the discrete memoryless half-duplex MARC and the corresponding additive white Gaussian noise channel. This scheme is developed based on the generalization of the Quantize-and-Forward (QF) scheme and single block with two slots coding structure. 2) as the general performance measure of the slow fading channel, the common outage probability and the expected sum rate (total throughput) of the GQF scheme have been characterized. The numerical examples show that when the relay has no access to the CSI of the relay-destination link, the GQF scheme outperforms other relaying schemes, e.g., classic compress-and-forward (CF), decode-and-forward (DF) and amplify-and-forward (AF). 3) for a MAC channel with heterogeneous user channels and quality-of-service (QoS) requirements, individual outage probability and total throughput of the GQF scheme are also obtained and shown to outperform the classic CF scheme.
\end{abstract}

\section{Introduction}
Although the general capacity of a static relay channel is still unknown, it is a well known fact that relaying can benefit a conventional point-to-point communication channel by cooperating with the transmitter \cite{Cover1979,Hodtani2009}. Moreover, it is also proved that the relaying can improve the sum achievable rates for a multiple access channel \cite{Kramer2005,Gunduz2010,Sahebalam2013}. The fundamental relaying schemes are based on decode-and-forward (DF) and compress-and-forward (CF). The CF based schemes are not limited by the decoding capability of the relay, and therefore they can be beneficial in cases where the relay is closer to the destination than to the source. Different variations of the CF based scheme have been investigated in \cite{Cover2007,Avestimehr2011,Lim2011,Razaghi2013,Wu2013}.

\begin{figure}[t]
\centering
\subfloat[MARC Slot-1]           {\label{fig:MARC-phase1}
\includegraphics[width=2.5in]{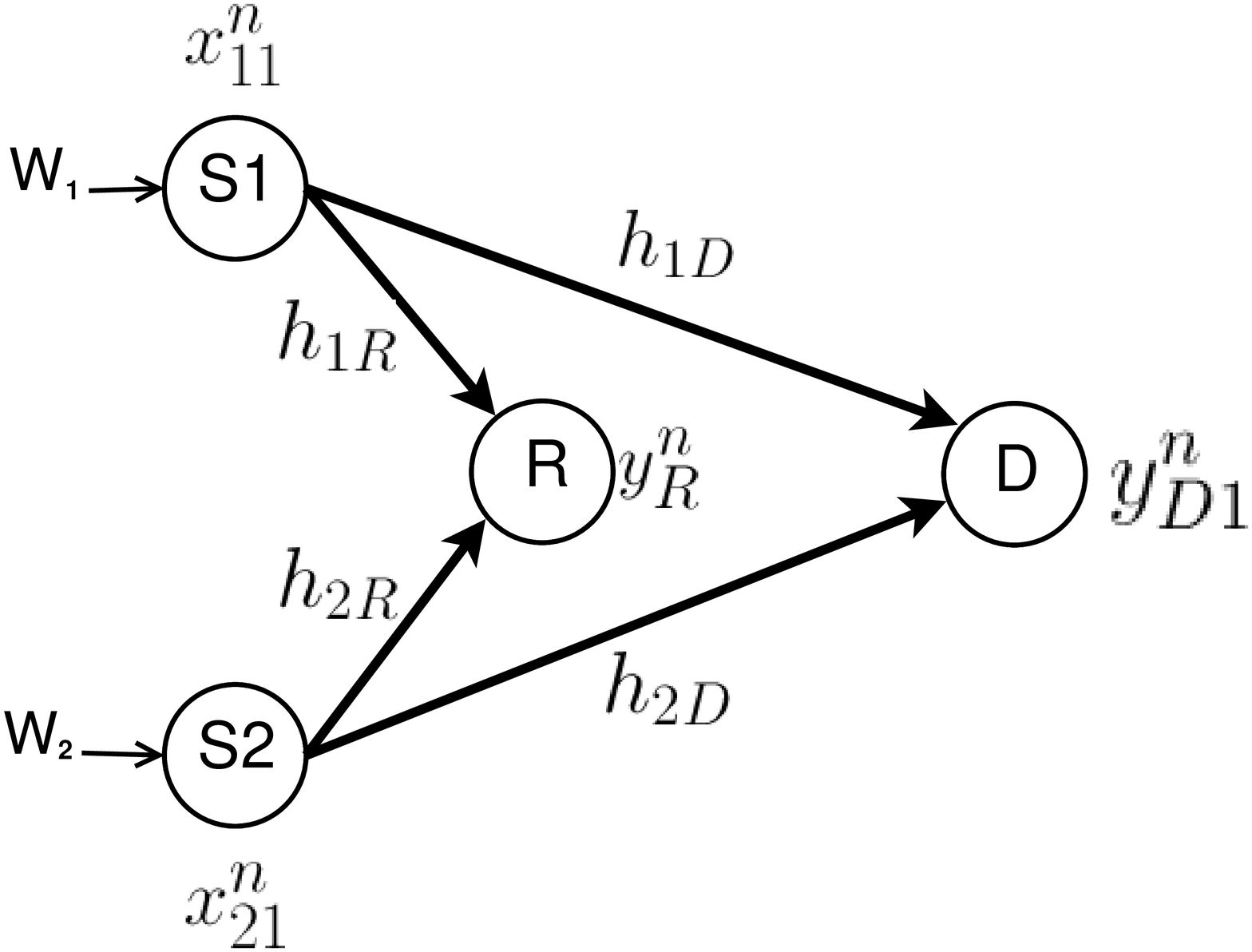}   }
\qquad
\subfloat[MARC Slot-2]{\label{fig:MARC-phase2}
\includegraphics[width=2.5in]{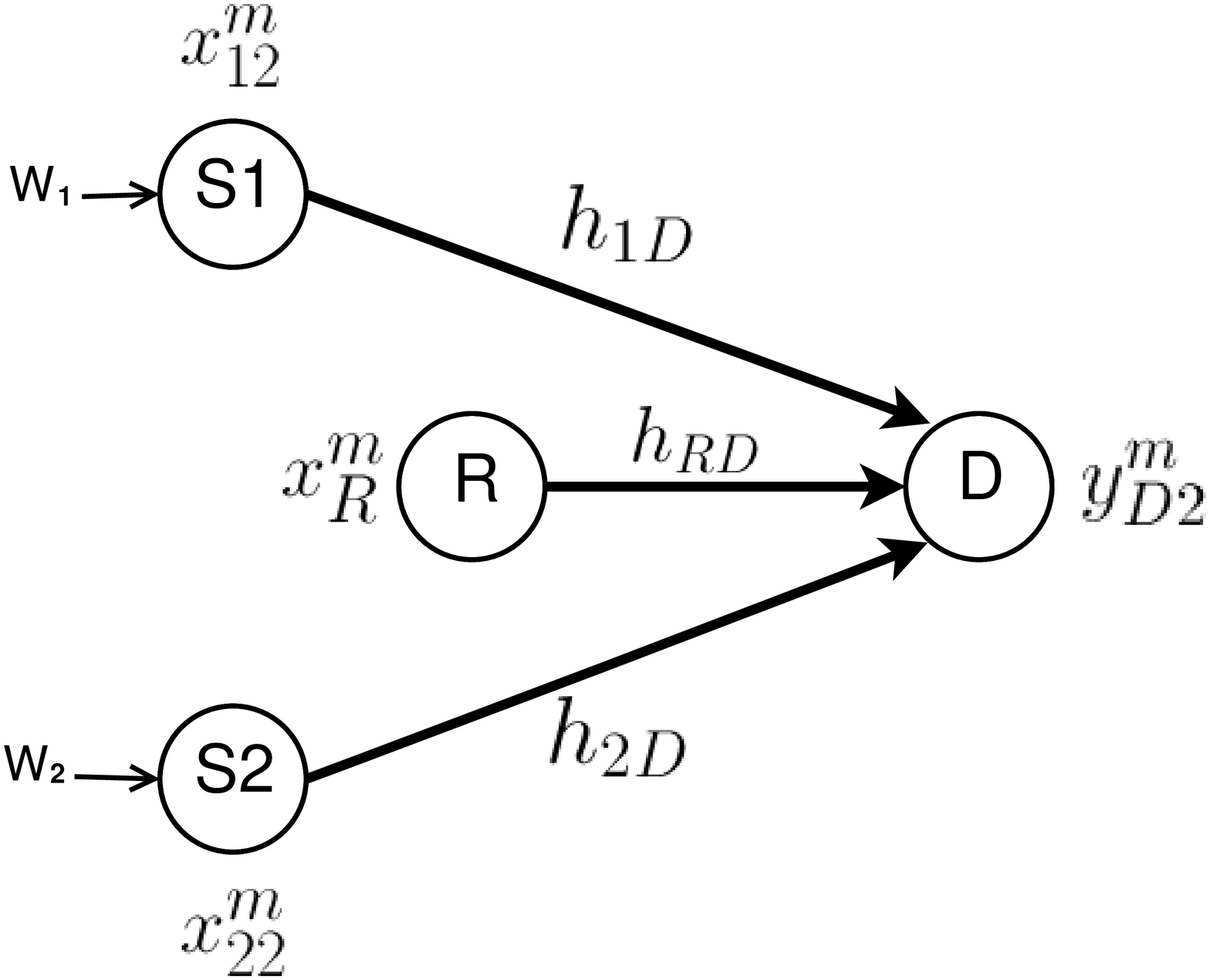}   }
\caption{Message flow of the HD-MARC.}
\label{fig:MARC-phases}
\end{figure}

In a slow fading wireless relay channel, the system outage probability can be decreased significantly by the diversity offered from the relaying \cite{Laneman2003,Chen2012}. Motivated by the practical constraint that relay cannot transmit and receive simultaneously in a wireless communication chanel\cite{Laneman2003,Khojastepour2003}, a slow fading Half-Duplex MARC (HD-MARC) (shown in Fig. \ref{fig:MARC-phases}) is considered in this paper. In particular, a block fading channel where the channel coefficients stay constant in each block but change independently from block to block is studied. In addition, it is assumed that the channel state information (CSI) is not available at the transmitter side. Specifically, the destination has complete CSI and the relay has only the CSI of the source-to-relay link.

The DF based schemes has been investigated over the slow fading MARC in \cite{Azarian2005}. The CF based schemes were shown to outperform the DF based schemes when the relay has the complete CSI of the channel\cite{Yuksel2007,Kim2009}. However, the perfect CSI at relay is generally too ideal. When the critical delay constraint exist in the wireless channels, the relay may not be able to obtain the CSI accurately. The CF-based schemes are not efficient in a slow fading environment when the relay has no access to the complete CSI \cite{Yao2013}. In \cite{Yao2013}, a quantize-and-forward (QF) scheme based on Noisy Network Coding (NNC) \cite{Lim2011} that applies a single block and two-slot coding structure, has been studied for a fading half-duplex relay channel (HDRC). In order to study the slow fading HD-MARC, a generalization of the QF scheme, the generalized quantized-and-forward (GQF), is proposed. Compared with the QF scheme, the proposed GQF scheme not only adopts the single block two slots coding structure but also takes into account the effect of the interference caused by the other message at the relay. Unlike the classic CF scheme, the GQF scheme only requires a simplified relay in the sense that no Wyner-Ziv binning is necessary. Also in the destination decoder, the GQF scheme uses the joint decoding instead of sequential decoding. As shown in this work, when the relay has no access to the CSI of the relay-destination link, the GQF can regain the advantage of using the CF based schemes.

In this work, we have generalized the QF scheme to the HD-MARC. Unlike the 3-node relay channel in \cite{Yao2013}, the GQF scheme now takes into account of the multi-user interference presented at relay. Moreover, without CSI of relay-destination link available at relay, the GQF scheme performs almost as good as the CF scheme with perfect CSI at relay in the slow fading channel. From this point, the GQF scheme is more practical since the CSI of the ongoing channel is not always available. 

From the engineering perspective, the result from this work can be applied (but not limited to) in the following two systems: 1) a multiuser cellular system with a cheap cost relay. Applying the GQF scheme at relay, no need of the CSI of R-D at relay, the probability of outage is greatly reduced compared to other relaying schemes.  2) a wireless sensor network where the relay is helping the communication between the sensor nodes to the data sink. Adding the cheap cost relay and use GQF scheme can improve the fading performance thus reduce the energy consumed by the sensor nodes.

The main contributions and the paper contents are summarized in the following:
\begin{enumerate}
\item In order to study the GQF scheme over the slow fading HD-MARC, the static channel has been characterized first. The achievable rate regions of the discrete memoryless half-duplex MARC and the corresponding additive white Gaussian noise (AWGN) channel are established based on the GQF scheme and the classic CF scheme. The performance comparison between the GQF scheme and the CF scheme is also discussed and shown with numerical example. It is shown that the GQF scheme can provide similar achievable rates while only a simplified relay (no binning necessary) is required.
\item Based on the achievable rates, the common outage probability and expected sum rate of the GQF scheme are derived and compared to the classic CF scheme and other common relaying schemes ( AF\cite{Laneman2003,Chen2008} and DF\cite{Gong2010}). It is shown by the numerical examples that, without relay-to-destination CSI at relay, the GQF scheme outperform the other schemes and can regain a large portion of the benefit provided by the CF-based schemes (with perfect CSI at relay) over DF-based schemes in the selected topology.
\item In practice, each user in a MAC may have different quality-of-service (QoS) requirement\cite{Narasimhan2007}. Similarly, for a two-user MARC, the destination failing to decode one of the source messages may not affect the other user's QoS requirement (message decoded successfully by the destination). Therefore, the individual outage related performance of the GQF scheme has also been discussed in terms of the individual outage probability and expected sum rate. The numerical examples are given to demonstrate the differences between the common and individual outage as well as the advantage of the GQF scheme.
\end{enumerate}

\section{Preliminaries}
\subsection{Half-Duplex Multiple Access Relay Channel}
A two-user half-duplex multiple access relay channel is considered in this paper as shown in Fig. \ref{fig:MARC-phases}. In particular, two sources $S_{1}$ and $S_{2}$ wish to send information to one destination $D$ with the help a relay $R$. Assume that each communication block length is $l$ channel uses and divided into two slots. The lengths of the first and the second slot are $n$ and $m$ channel uses, respectively. In the first slot, both $S_1$ and $S_2$ broadcast their messages to $R$ and $D$. In the second slot, $S_1$ and $S_2$ keep transmitting to $D$ while $R$ cooperates by transmitting to $D$ as well. Denote $x_{i1}^n$ and $x_{i2}^m$,  as the transmitted sequences by $S_i$ in the first and second slot correspondingly, and $x_{R}^m$ as the transmitted sequence by $R$ in the second slot, where $x_{ij}^k=[x_{ij,1},x_{ij,2},\cdots ,x_{ij,k}]$ and $x_{R}^k=[x_{R,1},x_{R,2},\cdots ,x_{R,k}]$ for $i,j \in \{1,2\}$ and $k \in \{n,m\}$. The received sequences at the destination in the first and the second slots are denoted as $y_{D1}^n$ and $y_{D2}^m$, respectively, and received sequence at the relay is $y_{R}^n$ in the first slot.

\subsection{Static Channels}

We consider two channel models. In the discrete memoryless channel case, the source output takes discrete values. In the Gaussian channel case, the source outputs are continuous values generated according to a Gaussian distribution. In both cases, the source is memoryless, i.e., the value of the source output at any given time is independent of the values at other times. These two static channel models are described as follows:

\subsubsection{Discrete Memoryless Channel}
In the discrete memoryless HD-MARC, all random variables take value from discrete alphabets. Each source $S_i$, $i=1,2$ chooses a message $W_i$ from a message set $\mathcal{W}_{i}=\{1,2,\dots,2^{lR_i}\}$, then encodes this message into a length $n$ codeword with an encoding function $f_{i1}(W_i)=X_{i1}^{n}$ and a length $m$ codeword with an encoding function $f_{i2}(W_i)=X_{i2}^{m}$, finally sends these two codewords in the corresponding slots. Relay $R$ employs an encoding function based on its reception $Y_{R}^{n}$ in the first slot.

Each destination uses a decoding function $g_{i}(Y_{i1}^{n},Y_{i2}^{m})=(\hat{W}_1,\hat{W}_2)$ that jointly decodes messages from the receptions in both slots. The channel transition probabilities can be represented by
\begin{eqnarray}
 p_{Y_{R}^{n}Y_{11}^{n}Y_{21}^{n}| X_{11}^{n}X_{21}^{n}}(y_R^n,y_{11}^n,y_{21}^n| x_{11}^n,x_{21}^n)
=\prod_{i=1}^{n}p_{Y_{R}Y_{11}Y_{21}| X_{11}X_{21}}(y_{R,i},y_{11,i},y_{21,i}| x_{11,i}x_{21,i})
\\
 p_{Y_{12}^{m}Y_{22}^{m}| X_{12}^{m}X_{22}^{m}X_{R}^{m}}(y_{12}^m,y_{22}^m| x_{12}^m,x_{22}^m,x_{R}^m)
 =\prod_{i=1}^{m}p_{Y_{12}Y_{22}| X_{12}X_{22}X_{R}}(y_{12,i},y_{22,i}| x_{12,i}x_{22,i}x_{R,i}).
\end{eqnarray}
for both slots. A rate pair $(R_1,R_2)$ is called achievable if there exists a message set, together with the encoding and decoding functions stated before such that $Pr(\hat{W}_1\neq W_1 \cup \hat{W}_2\neq W_2)\rightarrow 0$ when $l \rightarrow \infty $.

\subsubsection{Gaussian Channel}
In this case, the noise at each receiver is an additive white Gaussian (AWGN) random variable. The signal at each receiver is modeled as the faded transmitted signal corrupted by an AWGN component:
\begin{eqnarray}
y_{D1}^n & = & h_{1D}x_{11}^n+h_{2D}x_{21}^n+z_{D1}^n\\
y_{R}^n & = & h_{1R}x_{11}^n+h_{2R}x_{21}^n+z_{R}^n\\
y_{D2}^m & = & h_{1D}x_{12}^m+h_{2D}x_{22}^m+h_{RD}x_R^m+z_{D2}^m
\end{eqnarray}
where the channel coefficient $h_{ij}$ for $i\in\{1,2,R\}$ and $j\in\{R,D\}$ is real constant. The noise sequences of $z_{D1}^n, z_{D2}^m$ and $z_{R}^n$ are generated independently and identically with Gaussian distributions with zero means and unit variances.
In order to clarify the CF based relaying schemes, define the auxiliary random variable $\hat{Y}_R$ as the quantized signal of relay's recetption $Y_R$, i.e., $\hat{Y}_R = Y_R+Z_Q$, where $Z_Q$ is the quantization noise and is an independent Gaussian random variable with zero mean and variance $\sigma_Q^2$. The transmitters have power constraints over the transmitted sequences in each slot as $\frac{1}{n}\sum_{i=1}^{n}|x_{j,i}|^2 \leq P_j$ for $j\in\{11,21\}$ and $\frac{1}{m}\sum_{i=1}^{m}|x_{k,i}|^2 \leq P_k$ for $k\in\{12,22,R\}$, where $|x|$ shows the absolute value of $x$.

\subsection{Fading Channels}
Follows the similar notation of \cite{Yao2013}, denote the channel coefficient vector
\begin{equation}
\mathbf{h}:=[h_{1D},h_{2D},h_{1R},h_{2R},h_{RD}]\label{eqn-channelVec}.
\end{equation}
In this work, a block Rayleigh fading model is considered. Therefore, all the elements of $\mathbf{h}$ are assumed to be mutually independent and circularly symmetric complex Gaussian with zero means and variances $\sigma_{ij}^2$. They are constants within each block but change independently over different blocks. The noise sequences $z_{11}^n, z_{12}^m$ and $z_{R}^n$ are also circularly symmetric complex Gaussian with zero means and unit variances. Motivated by the practical applications, the source nodes have no CSI, i.e. no knowledge of $\mathbf{h}$. Hence, each of them can only use a coding scheme with fixed rate $R_i , i \in \{1,2\}$ to send messages. The relay has only receiver side CSI meaning only $h_{1R}$ and $h_{2R}$ are available. The destination knows $\mathbf{h}$ and therefore has complete CSI.
\begin{figure}[t]
\centering
\includegraphics[scale=0.3]{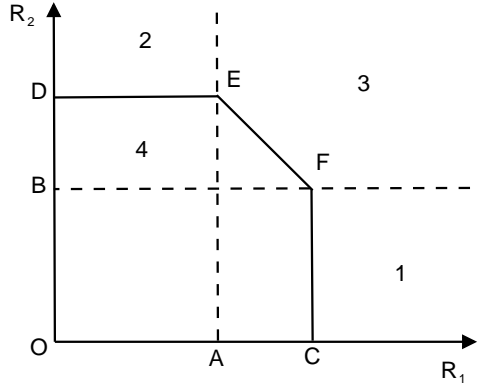}
\caption{Achievable rate region of a 2-user MARC conditioned on channel state}
\label{fig:mac}
\end{figure}

In Fig. \ref{fig:mac}, the achievable rate regions of a two-user MARC conditioned on the channel state is shown as the region 4 (bounded by two axis and the points C,D,E,F). The instantaneous achievable rate of a certain relaying scheme within a block of transmission is described as $I_{W,j}(\mathbf{h})$, where $j \in \{1,2,sum\}$ and $W$ denotes the relaying scheme. $I_{W,j}(\mathbf{h})$ is fixed for each block but a random entity determined by $\mathbf{h}$ within the entire transmission. In the following, the outage and the region of probabilities are taken over the random vector $\mathbf{h}$.

\subsubsection{Common outage probability and expected rate}
Similarly as \cite{Li2005,Narasimhan2007}, the common outage probability is defined as the probability that the chosen fixed rate pair $(R_1,R_2)$ lies outside the achievable rate region, given $\mathbf{h}$:
\begin{eqnarray}
P_{out,common}(R_1,R_2)=  Pr \{R_1+R_2> I_{W,sum}(\mathbf{h}) \;\;\text{or}\;
 R_1> I_{W,1}(\mathbf{h}) \;\;\text{or} \;\; R_2> I_{W,2}(\mathbf{h}) \}\label{eqn-common-outage}.
\end{eqnarray}
The system throughput or the expected sum rate is defined as in \cite{Yao2013} and \cite{Narasimhan2007}:
\begin{equation}
\bar{R}_{common}(R_1,R_2)=(R_1+R_2)(1-P_{out,common}(R_1,R_2)).
\end{equation}
The common expected rates can be obtained once the common outage probability is determined. Hence, section \MakeUppercase{\romannumeral 3} will focus on the common outage probabilities.

\subsubsection{Individual Outage of the MARC}
The individual outage event for a given user, say $S_1$, is defined as the message $W_1$  can not be decoded correctly at the destination, irrespective the successful decoding of the message $W_2$. According to Fig. \ref{fig:mac}, the common and individual outage probabilities can be described as:
{\setlength\arraycolsep{0.1em}
\begin{eqnarray}
P_{out,indiv1}(R_1,R_2)&=&P_{reg,1}+P_{reg,3}
\label{eqn-out-indiv1} \\
P_{out,indiv2}(R_1,R_2)&=&P_{reg,2}+P_{reg,3}
\label{eqn-out-indiv2} \\
P_{out,common}(R_1,R_2)&=&1-P_{reg,4}
=P_{reg,1}+P_{reg,2}+P_{reg,3}
\label{eqn-out-mac-common}
\end{eqnarray}
}
where $P_{reg,i}$, $i=1,2,3,4$ denotes the probability of the rate pair $(R_1, R_2)$ lies within the region $i$. Then the expected sum rate based on the individual outage probability is
{\setlength\arraycolsep{0.1em}
\begin{eqnarray}
\bar{R}_{indiv}(R_1,R_2)&=&R_1(1-P_{out,indiv1}(R_1,R_2))
+R_2(1-P_{out,indiv2}(R_1,R_2)) \label{eqn-out-indiv-expect-rate}.
\end{eqnarray}
}
\section{Achievable Rates In The Static Channels }
To study the performance of the proposed GQF scheme over the slow fading channel, the achievable rates for both discrete memoryless and the AWGN HD-MARC are derived.

\subsection{ Achievable rates in the Discrete-Memoryless Channel}
The GQF scheme is  an essential variation of the classic CF. In GQF, relay quantizes its observation $Y_R$ to obtain $\hat{Y}_R$ after the first slot, and then sends the quantization index $u\in\mathcal{U}=\{1,2, \cdots, 2^{lR_U}\}$ in the second slot with $X_R$. Unlike the conventional CF, no Wyner-Ziv binning is applied by the relay, which simplifies the relay operation. At the destination, decoding is also different in the sense that joint-decoding of the messages from both slots without explicitly decoding the quantization index is performed in GQF scheme.

The following theorem charactrize the achievable rate region for the discrete memoryless HD-MARC and its proof presents the detailed coding implementation of the GQF scheme:
\begin{theorem}\label{th-QFD-MARC}
The following rate regions are achievable over discrete memoryless HD-MARC based on the GQF scheme:
{\setlength\arraycolsep{0.01em}
\begin{eqnarray}
R_1 & < & \beta I(X_{11};Y_{D1},\hat{Y}_{R} | X_{21})
 + (1-\beta)I(X_{12};Y_{D2} | X_{22},X_{R})\label{eqn-GQF-R1}
\\
R_1 + R_U & < & \beta[I(X_{11},\hat{Y}_{R};Y_{D1} | X_{21})+I(X_{11};\hat{Y}_R)]
+ (1-\beta)I(X_{12},X_{R};Y_{D2} | X_{22})\label{eqn-GQF-R1RU}
\\
R_2 & < & \beta I(X_{21};Y_{D1},\hat{Y}_{R} | X_{11})
 +  (1-\beta)I(X_{22};Y_{D2} | X_{12}, X_{R})\label{eqn-GQF-R2}
\\
R_2 + R_U & < & \beta[I(X_{21},\hat{Y}_{R};Y_{D1} | X_{11})+I(X_{21};\hat{Y}_R)]
+ (1-\beta)I(X_{22},X_{R};Y_{D2} | X_{12})\label{eqn-GQF-R2RU}
\\
R_1 + R_2 & < & \beta I(X_{11},X_{21};Y_{D1},\hat{Y}_{R})
 + (1-\beta)I(X_{12},X_{22};Y_{D2} | X_{R})\label{eqn-GQF-R1R2}\\
R_1 + R_2 + R_U & < & \beta [I(X_{11},X_{21},\hat{Y}_{R};Y_{D1})+I(X_{11},X_{21};\hat{Y}_{R})]
\nonumber \\
&  & + (1-\beta)[I(X_{12},X_{22},X_{R};Y_{D2})\label{eqn-GQF-R1R2RU},
\end{eqnarray}
}
where  $\beta=n/l$ is fixed and
\begin{equation}
    R_U > \beta I(Y_R;\hat{Y}_R),\label{eq-rfindu}
\end{equation}
for all input distributions
\begin{equation}
p(x_{11})p(x_{21})p(x_{12})p(x_{22})p(x_R)p(\hat{y}_R|y_R).
\label{inputD-MARC}
\end{equation}
\end{theorem}
\begin{IEEEproof}:The detail of the proof can be found in the Appendix.
\end{IEEEproof}

\textit{Remark 1:} The major difference between the GQF scheme and the classic CF scheme applied in \cite{Gunduz2010} is that relay does not perform Wyner-Ziv binning after quantize its observation of the sources messages. Moreover, in GQF the destination performs one-step joint-decoding of both messages instead of sequentially decoding.

The achievable rate regions based on the classic CF scheme are shown for references.

\begin{theorem}\label{th-CFD-MARC}
The following rate regions are achievable over discrete memoryless half-duplex MARC based on the classic CF scheme:
{\setlength\arraycolsep{0.1em}
\begin{eqnarray}
R_1 & < & \beta I(X_{11};Y_{D1},\hat{Y}_{R} | X_{21})
+ (1-\beta)I(X_{12};Y_{D2} | X_{22},X_{R})\label{eqn-CF-R1}
\\
R_2 & < & \beta I(X_{21};Y_{D1},\hat{Y}_{R} | X_{11})
+  (1-\beta)I(X_{22};Y_{D2} | X_{12}, X_{R})\label{eqn-CF-R2}
\\
R_1 + R_2 & < & \beta I(X_{11},X_{21};Y_{D1},\hat{Y}_{R})
+ (1-\beta)I(X_{12},X_{22};Y_{D2} | X_{R}),\label{eqn-CF-R1R2}
\end{eqnarray}
}
subject to
\begin{equation}
\beta [I(Y_R;\hat{Y}_R)-I(Y_{D1};\hat{Y}_R)] <(1-\beta)I(X_R;Y_{D2})
\label{eqn-CFDcon-MARC}
\end{equation}
where  $\beta=n/l$ is fixed, for all the input distributions as in (\ref{inputD-MARC}).
\end{theorem}
\begin{IEEEproof}
Due to the similarity of CF and GQF schemes, the detailed proof is omitted. Note that the classic CF scheme has been modified to fit the HD MARC channel. Now the relay quantizes the received signal with rate $R_U$ after first slot, applies the Wyner-Ziv binning to further partition the set of alphabets $\mathcal{U}$ into $2^{lR_S}$ equal size bins and sends the bin index $S$ with $X_{R}(s)$ in the second slot. The destination performs successive decoding, i.e. sequentially decodes the bin index $\hat{s}\in \mathcal{S}$, quantization index $\hat{u} \in B(\hat{s})$ with the side information and finally the source messages $(\hat{w_1},\hat{w_2})\in (\mathcal{W}_1,\mathcal{W}_2)$ jointly from both slots reception.
\end{IEEEproof}

\textit{Remark 2:} The GQF and the classic CF schemes generally provide different achievable rate regions. Note that the achievable results (\ref{eqn-CF-R1})-(\ref{eqn-CF-R1R2}) should have (\ref{eqn-CFDcon-MARC}) hold, which means the relay-destination link is good enough to support the compression at relay to be recovered at destination. In theorem 1, the individual rate $R_1$ is determined by the minimum of \eqref{eqn-GQF-R1} and \eqref{eqn-GQF-R1RU}, where we have applied \eqref{eq-rfindu} into \eqref{eqn-GQF-R1RU}. Given \eqref{eqn-CFDcon-MARC} satisfied in theorem 1, the right-hand side of \eqref{eqn-GQF-R1RU} is greater than the right-hand side of \eqref{eqn-GQF-R1}. In such case, $R_1$ is only determined by \eqref{eqn-GQF-R1}. As \eqref{eqn-GQF-R1} is the same as \eqref{eqn-CF-R1R2}, the GQF scheme and the CF scheme lead to the same individual rate $R_1$. Similarly, both schemes also result the same rates $R_2$ and $R_1+R_2$. In other words, if (24) holds, both GQF and CF schemes have the same achievable rates. Therefore, when (\ref{eqn-CFDcon-MARC}) holds and a simplified relay is not required, either the CF or the GQF scheme can be applied in the HD-MARC. On the other hand, if a low-cost simplified relay is preferred or (\ref{eqn-CFDcon-MARC}) does not hold, the GQF scheme is a superior choice.

\subsection{Achievable rates in the Gaussian Channels}
When the relay node knows only the CSI of the source to relay (S-R) link in a gaussian half-duplex relay channel, extending the achievable rate results from the discrete memoryless channel to the gaussian channel are not straightforward as shown in \cite{Yao2013}. To overcome this problem, a discretization approach has been proposed in \cite{Yao2013}. In this paper, since the relay node has only the CSI of the two sources to relay link as well, the discretization approach is adopted to HD-MARC and applied to derive the achievable rates of the GQF scheme.
\begin{proposition}
The following rate regions are achievable for the Gaussian HD-MARC with the GQF scheme, for which the relay node has only the CSI of S-R link:
{\setlength\arraycolsep{0.1em}
\begin{eqnarray}
R_i &<&  \underset{\sigma_Q^2,\beta}{max} \;\; min \{\frac{\beta}{2}log(1+h_{iD}^2P_{i1}+\frac{h_{iR}^2P_{i1}}{1+\sigma_Q^2})
+\frac{1-\beta}{2}log(1+h_{iD}^2P_{i2}),
\nonumber \\
&&\frac{\beta}{2}log(\frac{(1+h_{iD}^2P_{i1})\sigma_Q^2}{1+\sigma_Q^2})
+\frac{1-\beta}{2}log(1+h_{iD}^2P_{i2}+h_{RD}^2P_R)\}
\label{eqn-GQF-GauMARCindi}\\
R_1 &+& R_2 <   \underset{\sigma_Q^2,\beta}{max} \;\; min \{\frac{\beta}{2}log(1+h_{1D}^2P_{11}+h_{2D}^2P_{21}
\nonumber \\
&&+\frac{(h_{1D}h_{2R}-h_{1R}h_{2D})^2P_{11}P_{21}+h_{1R}^2P_{11}+h_{2R}^2P_{21}}{1+\sigma_Q^2})
+\frac{1-\beta}{2}log(1+h_{1D}^2P_{12}+h_{2D}^2P_{22}),
\nonumber \\
&&\frac{\beta}{2}log(\frac{(1+h_{1D}^2P_{11}+h_{2D}^2P_{21})\sigma_Q^2}{1+\sigma_Q^2})
+\frac{1-\beta}{2}log(1+h_{1D}^2P_{12}+h_{2D}^2P_{22}+h_{RD}^2P_R)\}
\label{eqn-GQF-GauMARCsum}
\end{eqnarray}
}
where i = 1,2 and $\sigma_Q^2$ is the variance of the relay quantization noise.
\end{proposition}

Note that though a discretization approach from [17] was adopted and applied to the HD-MARC, \eqref{eqn-GQF-GauMARCindi} and \eqref{eqn-GQF-GauMARCsum} are the same as the rate regions which were derived directly from Theorem 1. Let $i=1$, the first and the second minimum term of \eqref{eqn-GQF-GauMARCindi} are derived from \eqref{eqn-GQF-R1} and \eqref{eqn-GQF-R1RU}, respectively. Similarly, when $i=2$, \eqref{eqn-GQF-R2} and \eqref{eqn-GQF-R2RU} result \eqref{eqn-GQF-GauMARCindi}. The sum rate constraint \eqref{eqn-GQF-GauMARCsum} are obtained from \eqref{eqn-GQF-R1R2} and \eqref{eqn-GQF-R1R2RU}.

\begin{figure}[t]
\centering
\includegraphics[scale=0.25]{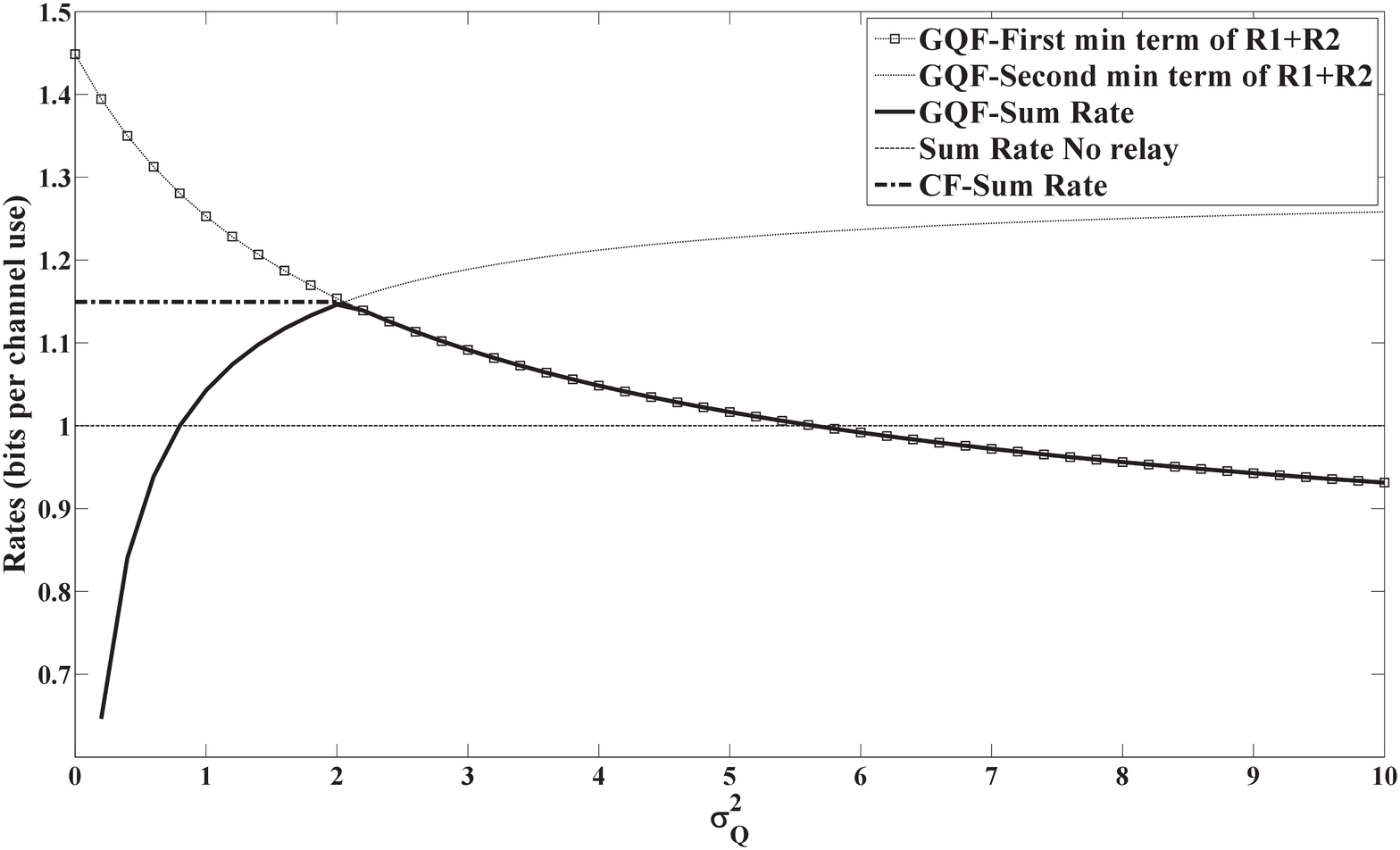}
\caption{Achievable Rates of GQF and CF based scheme with variant $\sigma_Q^2$, $h_{11}=h_{21}=1, h_{1R}=3, h_{2R}=0.5, h_{R1}=3,\beta=0.5$, $P_i=1$, where $i\in\{11,12,21,22,R\}$. For the case of no relay, sources have power $P_1=P_2=1.5$}
\label{fig:Sigma-q-GQF}
\end{figure}

\emph{Remark 3}:\; Similarly as \cite{Yao2013}, within the achievable sum rate (\ref{eqn-GQF-GauMARCsum}), the two min terms are two functions of $\sigma_Q^2$, i.e. $I_1(\sigma_Q^2)$ and $I_2(\sigma_Q^2)$, respectively. The impact of the different values of the $\sigma_Q^2$ on the achievable sum rate is shown in the Fig. \ref{fig:Sigma-q-GQF}. It can be seen that, for fixed $\beta$, $I_1(\sigma_Q^2)$ is a monotonically decreasing function and  $I_2(\sigma_Q^2)$ is a monotonically increasing function. Let $I_1(\sigma_Q^2)=I_2(\sigma_Q^2)$, the $\sigma_Q^2$ that maximizes the sum rate can be obtained.

Assuming $\beta$ is fixed, different values of $\sigma_Q^2$ can be selected to maximize the sum rate (\ref{eqn-GQF-GauMARCsum}) or the individual rate (\ref{eqn-GQF-GauMARCindi})
\begin{equation}
\sigma_{Qsum}^2 =
\frac{1+\frac{h_{1R}^2P_{11}+h_{2R}^2P_{21}+(h_{1D}h_{2R}-h_{1R}h_{2D})^2P_{11}P_{21}}{1+h_{1D}^2P_{11}+h_{2D}^2P_{21}}}{(1+\frac{h_{RD}^2P_R}{1+h_{1D}^2P_{12}+h_{2D}^2P_{22}})^{\frac{1-\beta}{\beta}}-1},
\label{eqn-GQF-GauMaxSum}
\end{equation}
\begin{equation}
\sigma_{Qindiv,i}^2 =
\frac{1+\frac{h_{iR}^2P_{i1}}{1+h_{iD}^2P_{i1}}}{(1+\frac{h_{RD}^2P_R}{1+h_{iD}^2P_{i2}})^{\frac{1-\beta}{\beta}}-1},
\label{eqn-GQF-GauMaxIndiv}
\end{equation}
where $i=\{1,2\}$. For the static MARC,  sum rate is a more important rate constraint term than the individual one since it captures the overall performance the multi-user channel. Therefore, $\sigma_Q^2$ is usually chosen to maximize (\ref{eqn-GQF-GauMARCsum}). However, in the slow fading MARC, when relay has complete CSI and sometimes choosing $\sigma_{Qindiv,i}^2$ instead of $\sigma_{Qsum}^2$ can reduce the common outage probability of the GQF scheme.

As reference, the achievable rates based on the classic CF scheme is shown below:
\begin{proposition}
The following rates are achievable for the Gaussian HD-MARC by using the classic CF scheme:
{\setlength\arraycolsep{0.1em}
\begin{eqnarray}
R_i&<&\frac{\beta}{2}log(1+h_{iD}^2P_{i1}+\frac{h_{iR}^2P_{i1}}{1+\sigma_Q^2})
+\frac{1-\beta}{2}log(1+h_{iD}^2P_{i2}),
\label{eqn-CF-GauMARCindi}
\\
R_1 + R_2 &<& \frac{\beta}{2}log(1+h_{1D}^2P_{11}+h_{2D}^2P_{21}
+\frac{(h_{1D}h_{2R}-h_{1R}h_{2D})^2P_{11}P_{21}+h_{1R}^2P_{11}+h_{2R}^2P_{21}}{1+\sigma_Q^2})
\nonumber \\
&+&\frac{1-\beta}{2}log(1+h_{1D}^2P_{12}+h_{2D}^2P_{22})
\label{eqn-CF-GauMARCsum}
\end{eqnarray}
}
where i=1,2 and
\begin{equation}
\sigma_Q^2 >
\frac{1+\frac{h_{1R}^2P_{11}+h_{2R}^2P_{21}+(h_{1D}h_{2R}-h_{1R}h_{2D})^2P_{11}P_{21}}{1+h_{1D}^2P_{11}+h_{2D}^2P_{21}}}{(1+\frac{h_{R1}^2P_R}{1+h_{1D}^2P_{12}+h_{2D}^2P_{22}})^{\frac{1-\beta}{\beta}}-1}.
\label{eqn-CF-GauMARC}
\end{equation}
\end{proposition}
\emph{Remark 4}:\; The sum rate (\ref{eqn-CF-GauMARCsum}) is the same as the first min term of (\ref{eqn-GQF-GauMARCsum}) when (\ref{eqn-CF-GauMARC}) is satisfied. Fig. \ref{fig:Sigma-q-GQF} also shows (\ref{eqn-CF-GauMARCsum}) with different $\sigma_Q^2$. (\ref{eqn-CF-GauMARC}) is the condition that makes CF scheme work. A smaller value of $\sigma_Q^2$ means $\hat{Y}_R$ is a less compressed observation of $Y_R$, and hence a higher rate of $\hat{Y}_R$. On the other hand, the channel between relay and destination requires the rate of $\hat{Y}_R$ to be small enough  since the compression should be recovered by the destination.

Notice that if relay uses a good quantizer or relay has a good estimate $\hat{Y}_R$ of $Y_R$ such that $\sigma_Q^2$ is less than the right-hand side of (\ref{eqn-CF-GauMARC}). Then, a higher sum rate cannot be achieved. This is due to the channel between relay and destination is limiting the compressed observation at relay to be recovered at destination. In other words, $\hat{Y}_R$ has a higher rate than the channel between relay and destination can support. Thus for smaller value of $\sigma_Q^2$, the sum rate is the same as the one taken from the constraint condition.

\begin{figure}[t]
\centering
\includegraphics[scale=0.25]{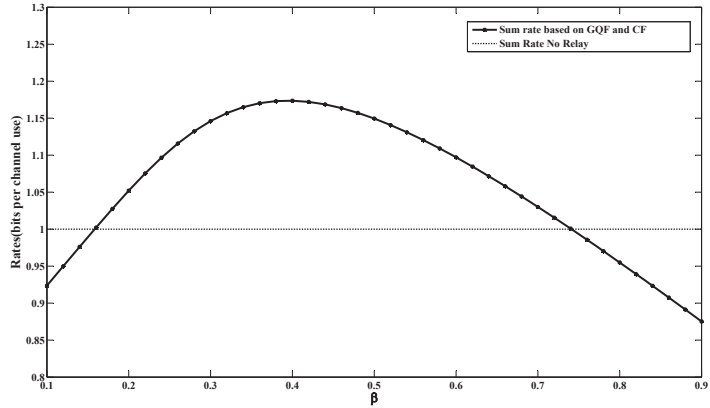}
\caption{Achievable Rates of GQF and CF based scheme with variant $\beta$}
\label{fig:beta-factor}
\end{figure}

As defined previously, $\beta$ is the ratio of the two slots taken in each block. The impact of the factor $\beta$ on the achievable rates that based on the GQF scheme in the HD-MARC channel is shown in Fig. \ref{fig:beta-factor} where we assume the same power and channel gain as Fig. \ref{fig:Sigma-q-GQF}. It can be seen that under such a channel state in order to maximize the achievable sum rate the length of each slot should be carefully chosen. Notice that if the relay quantization random variable $\sigma_Q^2$ was chosen to satisfy the constraint (\ref{eqn-CF-GauMARC}), and then the achievable sum rates based on the modified classic CF is the same as those based on GQF. As also shown in the Fig. \ref{fig:beta-factor}, both GQF and CF schemes outperform the case in which no relay is available in the channel.

The achievable rates of the GQF scheme with optimized $\sigma_Q^2$ are the same as the classic CF scheme over Gaussian channels. However, as shown in section \MakeUppercase{\romannumeral 3} and \MakeUppercase{\romannumeral 4}, the GQF scheme is able to provide significant gain over the CF scheme in the slow fading HD-MARC when the relay node has only receiver CSI.

\section{Common Outage Related Performance}
Based on the achievable rate region result from previous section, the common outage events and outage probability of the GQF scheme are characterized in this section.

\subsection{Outage Probability of the GQF scheme}

 Since the source nodes have no access to the CSI, $S_1$ and $S_2$ can only use a fixed rate pair of $(R_1,R_2)$ to transmit. The relay node has no CSI of the R-D link, therefore it is not able to adapt to the channel state $\mathbf{h}$ and choose rate $R_U$ accordingly. Instead, the relay can only use a fixed rate of $R_U$. In order to do so, the relay chooses the auxiliary random variable $\hat{Y}_R$ according to $\hat{Y}_R=Y_R+Z_Q$. The variance of the $Z_Q$ is chosen to have
\begin{equation}
   R_U =\beta I(Y_R;\hat{Y}_R)=\beta log(1+\frac{1+h_{1R}^2P_{11}+h_{2R}^2P_{21}}{\sigma_Q^2}). \label{eqn-fading-GQF-Ru}
\end{equation}
As every parameter in (\ref{eqn-fading-GQF-Ru}) is known, the relay can choose such $\sigma_Q^2$ successfully.

The GQF scheme employs the joint-decoding technique at the destination node, thus the common outage event happens when either one of the conditions (\ref{eqn-GQF-R1})-(\ref{eqn-GQF-R1R2RU}) not satisfied. Define the following sets:
{\setlength\arraycolsep{0.1em}
\begin{eqnarray}
&\mathcal{O}_{R_1} &:= \{\mathbf{h}:R_1 > \beta I(X_{11};Y_{D1},\hat{Y}_{R} | X_{21})
+ (1-\beta)I(X_{12};Y_{D2} | X_{22},X_{R})\}
\label{eqn-out1}\\
&\mathcal{O}_{R_{1u}} &:= \{\mathbf{h}:R_1>\beta[I(X_{11},\hat{Y}_{R};X_{21},Y_{D1})+I(X_{11};\hat{Y}_R)]
\nonumber \\
&&+ (1-\beta)I(X_{12},X_{R};Y_{D2} | X_{22})-R_U\}
\label{eqn-out2}\\
&\mathcal{O}_{R_2} &:= \{\mathbf{h}:R_2 > \beta I(X_{21};Y_{D1},\hat{Y}_{R} | X_{11})
+  (1-\beta)I(X_{22};Y_{D2} | X_{12}, X_{R})\}
\label{eqn-out3}\\
&\mathcal{O}_{R_{2u}} &:= \{\mathbf{h}:R_2> \beta[I(X_{21},\hat{Y}_{R};Y_{D1} | X_{11})+I(X_{21};\hat{Y}_R)]
\nonumber \\
&&+ (1-\beta)I(X_{22},X_{R};Y_{D2} | X_{12})-R_U\}
\label{eqn-out4}\\
&\mathcal{O}_{R_{12}} &:= \{\mathbf{h}:R_1 + R_2 >\beta I(X_{11},X_{21};Y_{D1},\hat{Y}_{R})
+ (1-\beta)I(X_{12},X_{22};Y_{D2} | X_{R})\}
\label{eqn-out5}\\
&\mathcal{O}_{R_{12u}} &:= \{\mathbf{h}:R_1 + R_2 >\beta [I(X_{11},X_{21},\hat{Y}_{R};Y_{D1})
\nonumber \\
&&+I(X_{11},X_{21};\hat{Y}_{R})] + (1-\beta)[I(X_{12},X_{22},X_{R};Y_{D2})-R_U\}\label{eqn-out6}
\end{eqnarray}
}
Since in (\ref{eqn-fading-GQF-Ru}), $R_U$ has been chosen to satisfy (\ref{eq-rfindu}), the common outage event is determined by the aforementioned six sets. The common outage probability of the GQF scheme is:
\begin{equation}
P_{out,common}^{GQF}(R_1,R_2,R_U)=Pr\{\mathcal{O}_{R_1} \cup \mathcal{O}_{R_{1u}} \cup \mathcal{O}_{R_2} \cup \mathcal{O}_{R_{2u}} \cup \mathcal{O}_{R_{12}} \cup \mathcal{O}_{R_{12u}}  \}\label{eqn-out-common}.
\end{equation}

On the other hand, if the relay node has the access to the complete CSI (knows $\mathbf{h}$), it can adjust $R_U$ according to that specific channel state. The common outage probabilities of the GQF scheme and the CF scheme are the same in this case and can be shown as:
\begin{equation}
P_{out}^{CSIT}=Pr\{R_1+R_2>I_{GQF,sum}(\mathbf{h})\;\;\text{or}\;\;
R_1>I_{GQF,1}(\mathbf{h}) \;\;\text{or}\;\; R_2>I_{GQF,2}(\mathbf{h})\}
\end{equation}
where $I_{GQF,i}(\mathbf{h})$, $i=1,2$  and $I_{GQF,sum}(\mathbf{h})$ are the right-hand sides of \eqref{eqn-GQF-GauMARCindi} and \eqref{eqn-GQF-GauMARCsum}, respectively.

\subsection{Numerical Examples and Discussions}
\begin{figure}[htpb]
\centering
\includegraphics[scale=0.45]{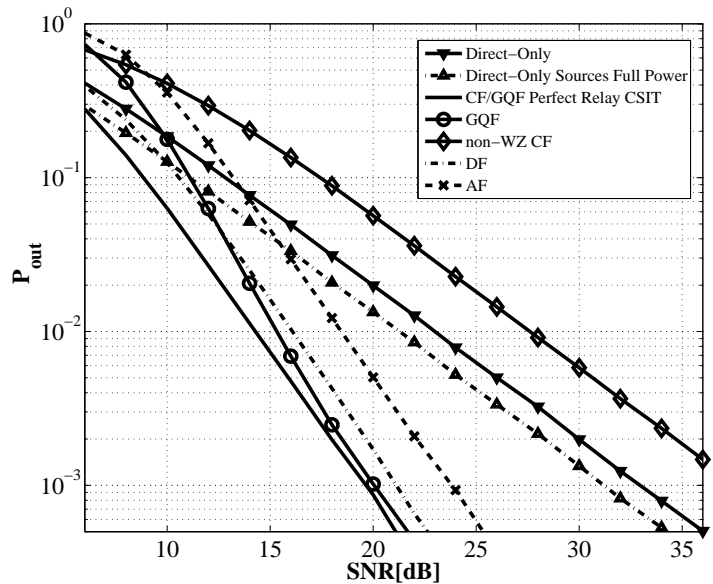}
\caption{Outage Probability of the GQF scheme, where $R_1=R_2=1$, $\beta=0.5$ and $R_U=3$. The outage probabilities of the CF, DF, AF and Direct-Transmission are shown for comparison}
\label{fig:outage1}
\end{figure}

\begin{figure}[htpb]
\centering
\includegraphics[scale=0.45]{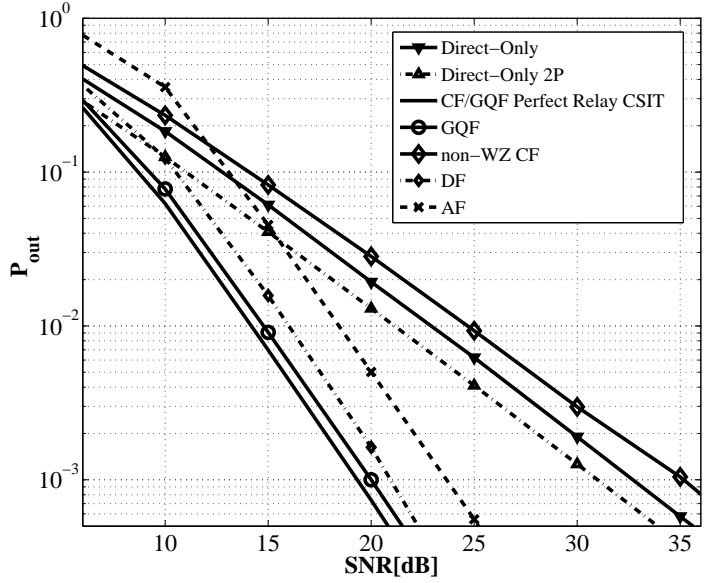}
\caption{Outage Probabilities of the GQF scheme and other schemes, where $R_1=R_2=1$,$\beta=0.5$ and $R_U$ is optimized for GQF and CF scheme}
\label{fig:outage2}
\end{figure}

Two examples of the common outage probabilities as functions of SNR are shown in Fig. \ref{fig:outage1} and Fig. \ref{fig:outage2}, where $\beta=0.5$, sources choosing fixed rate of $R_1=R_2=1$ bit/channel use and the variances of the channel coefficients in (\ref{eqn-channelVec}) $\sigma_i^2=1$ for $i \in \{1D,2D,1R,2R,RD\}$. Furthermore, with the power assumptions $P_{11}=P_{21}=P_{12}=P_{22}=SNR$ and $P_R=SNR/(1-\beta)$, the source nodes and the relay nodes has the same average power in each block of transmission. The direct transmission scheme is also presented as a reference. In the direct transmission scheme the relay is assumed to be silent during the whole block transmission, therefore the system is equivalent to a 2-user half-duplex MAC. The common outage probability of the direct transmission can be described similarly as (\ref{eqn-common-outage}).
The direct transmission scheme where each of the source node has 1.5 times power is also presented, which considers the case that the relay keeps silent in the whole block and does not consume any power. Hence, the overall power consumed by this direct scheme is the same as other schemes in which the relay transmit in the second block. The outage probabilities of the DF\cite{Gong2010}, AF \cite{Chen2008} and non-Wyner-Ziv (non-WZ) CF \cite{Kim2009} scheme are also shown in the examples for comparison.
The non-WZ CF \cite{Kim2009} is a special type of classic CF scheme. Such scheme employs the successful decoding at the destination but no Wyner-Ziv random binning at the relay, which simplifies the relay. The non-WZ CF scheme has no worse outage probability performance than the classic CF scheme as shown in \cite{Yao2013}. For simplification purposes, the non-WZ CF is used for comparing with the GQF in the numerical examples.

The outage probabilities of the above schemes are shown in Fig.\ref{fig:outage1}. It can be seen that without the R-D link CSI at relay, the non-WZ CF scheme performs significantly worse than all the other relaying schemes as well as the direct transmission scheme. Notice that the non-WZ CF scheme even does not provide any diversity gain. On the other hand, the GQF, DF and AF schemes show the diversity advantages of relaying. The GQF scheme with fixed $R_U=3$ performs very close to the GQF/CF scheme with complete CSI at relay. Comparing with the AF scheme, the GQF scheme outperforms in all SNR regions. The DF scheme has a smaller value of outage probability in low SNR regions comparing to GQF scheme. However, in higher SNR regions, the GQF scheme outperforms DF scheme.

In Fig. \ref{fig:outage2}, the same outage probabilities for different schemes are shown except that the GQF scheme is now with optimized $R_U$. The GQF scheme outperforms DF scheme in all SNR regions. Without perfect CSI at relay, even with the optimized $R_U$, the non-WZ CF scheme is still inefficient. It is due to the successive decoding (Sequential decoding) applied at the destination. In non-WZ CF scheme, the destination tries to decode the bin index sent by the relay first, then the compression index with the side information and the sources messages in the last. If the destination is not able to recover the bin index, it tries to decode the source messages while treating the signal from the relay $X_R$ as interference. Hence, with perfect CSI, the relay can adapt to the channel and chooses the $R_U$ correspondingly. However, without the perfect CSI, using non-WZ CF scheme become inefficient comparing to the joint-decoding based GQF scheme.

\begin{figure}[htpb]
\centering
\includegraphics[scale=0.4]{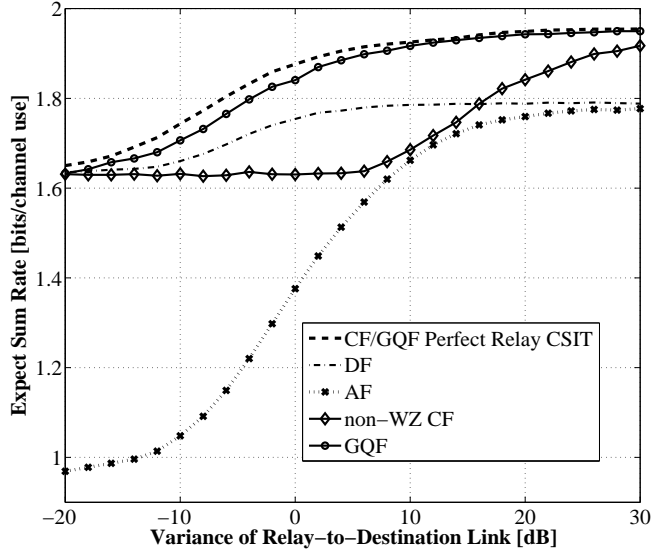}
\caption{Expected Rates of the different schemes}
\label{fig:exprate1}
\end{figure}

Only taking the common outage probability as the slow fading performance measure has some limitations. The result can be affected by the fixed rate pair of $(R_1,R_2)$. In order to gain further insight on the performance of those relaying schemes, the common expected rate is discussed. Assuming $\beta=0.5$, $P_{11}=P_{12}=P_{12}=P_{21}=SNR=10 dB$, $P_R=SNR/(1-\beta)$ and $\sigma_i^2=1$ for $i=1D,2D,1R,2R$, the common expected rate for the case where the relay is moving towards the destination by increasing the variance of the R-D link $\sigma_{RD}^2$ is obtained in Fig. \ref{fig:exprate1}. It can be seen that for the smaller $\sigma_{RD}^2$, the GQF and non-WZ CF schemes converge to the line which describes the direct transmission scheme.
For the larger $\sigma_{RD}^2$ (relay closes to destination), the GQF and non-WZ CF schemes outperform DF and AF schemes. The reason is that if the relay is close to the destination, the CF-based schemes (including GQF, classic Wyner-ZIv CF, etc) have better performance than the DF-based schemes \cite{Kramer2005,Gunduz2010}. Furthermore, the GQF scheme outperforms non-WZ CF scheme and perform close to the case GQF/CF with complete CSI.

\section{Individual Outage Related Performance}
In this section, the individual outage probabilities and the expected sum rates of the GQF scheme are characterized.

\subsection{Individual Outage of GQF scheme}
The individual outage event and outage probability of the GQF scheme are charcterized in this subsection. Specifically, $P_{out,indiv1}^{GQF}(R_1,R_2,R_U)$  of the source $S_1$ is derived. The $P_{out,indiv2}^{GQF}(R_1,R_2,R_U)$ for the source $S_2$ can be obtained similarly. Note that since no relay-to-destination CSI available at relay, it choses a fixed rate $R_U$ to transmit in the second slot. Different choices of $R_U$ will have different impacts on the individual outage probabilities, which is the same case as the common outage probability in previous section.

From (\ref{eqn-out-indiv1}) and (\ref{eqn-out-mac-common}), $P_{out,indiv1}^{GQF}(R_1,R_2,R_U)$ can be found by
\begin{eqnarray}
P_{out,indiv1}^{GQF}(R_1,R_2,R_U)=P_{reg,1}+P_{reg,3}
 =P_{out,common}^{GQF}(R_1,R_2,R_U)-P_{reg,2}
\label{eqn-out-com-indiv1}.
\end{eqnarray}

Therefore, only $P_{reg,2}$ is needed to obtain $P_{out,indiv1}^{GQF}$ as $P_{out,common}^{GQF}(R_1,R_2,R_U)$ is known from (\ref{eqn-out-common}). Given the channel fading state $\mathbf{h}$, region 2 in Fig. \ref{fig:mac} can be characterized by the following two conditions: 1) the decoder can decode the message $W_1$ successfully while treating the signals of $W_2$ as interference; 2) the decoder can not decode $W_2$ even with the successful interference cancelation of $W_2$, hence $W_2$ is in outage. Define the following sets:
{\setlength\arraycolsep{0.1em}
\begin{eqnarray}
&\mathcal{O}_{R_1,indiv1,1} &:= \{\mathbf{h}:R_1 > \beta I(X_{11};Y_{D1},\hat{Y}_{R})+ (1-\beta)I(X_{12};Y_{D2} |X_{R})\}
\\
&\mathcal{O}_{R_1,indiv1,2} &:= \{\mathbf{h}:R_1>\beta[I(X_{11},\hat{Y}_{R};Y_{D1})+I(X_{11};\hat{Y}_R)]
\nonumber \\
&&+ (1-\beta)I(X_{12},X_{R};Y_{D2} )-R_U\}.
\end{eqnarray}
}Then the condition 1) corresponds to $\mathcal{O}_{R_1,indiv1,1}^{c}$ and $\mathcal{O}_{R_1,indiv1,2}^{c}$, where $\mathcal{O}^{c}$ defines a complement set of $\mathcal{O}$. The previously defined sets $\mathcal{O}_{R_2}$ in (\ref{eqn-out3}) and $\mathcal{O}_{R_{2u}}$ in  (\ref{eqn-out4}) describe the condition 2). Thus, $P_{reg,2}$ is calculated as
\begin{equation}
P_{reg,2}= Pr(\mathcal{O}_{R_1,indiv1,1}^{c}     \cap  \mathcal{O}_{R_1,indiv1, 2}^{c}  \cap  \mathcal{O}_{R_2}  \cap  \mathcal{O}_{R_{2u}}).
\end{equation}

$P_{out,indiv1}^{GQF}(R_1,R_2,R_U)$ is then obtained by (\ref{eqn-out-com-indiv1}). The individual outage probability for $S_2$, $P_{out,indiv2}^{GQF}(R_1,R_2,R_U)$,  can be derived in a similar fashion. Applying (\ref{eqn-out-indiv-expect-rate}), the expected sum rate of the GQF scheme based on the individual outage probability is characterized.

\subsection{Numerical Examples}

\begin{figure}[t]
\centering
\subfloat[Individual outage probabilities]
{\label{fig:outage3}
\includegraphics[width=3.2in,height=3.2in]{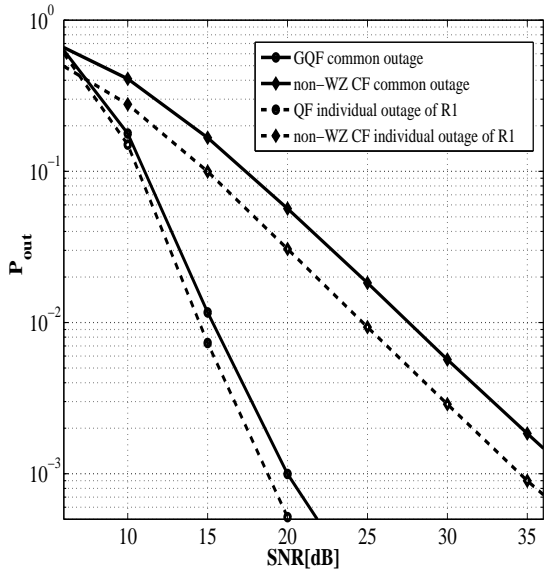}   }
\subfloat[Expected Rates] {\label{fig:exprate2}
\includegraphics[width=3.2in,height=3.2in]{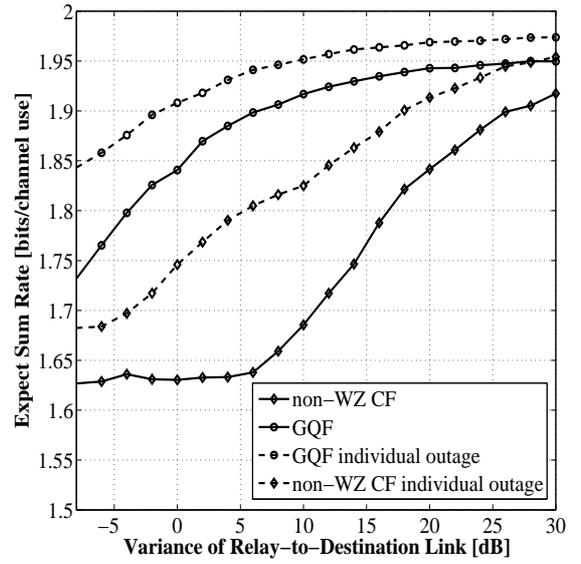}   }
\caption{Individual outage related performance}
\label{fig:indiv}
\end{figure}

The numerical examples of the individual outage performance are shown in Fig. \ref{fig:indiv}. The simulation parameters are the same as those in Section IV. It can be seen from Fig. \ref{fig:outage3} that the individual outage probabilities of the GQF and CF scheme are indeed smaller than the respective common outage probabilities. Fig. \ref{fig:exprate2} shows the total throughput of the GQF and CF scheme with individual outage probabilities are significantly higher than those with common outage probabilities. In addition, the GQF scheme outperforms the CF scheme in largeer $\sigma_{RD}^2$, which once again shows the advantage of using GQF scheme in slow fading MARC where no CSI of the R-D link available at relay.

\section{Conclusion}
In this paper, the GQF scheme in the slow fading half-duplex Multiple Access Relay Channel has been studied. First, the achievable rate regions were obtained for the discrete memoryless channel and the AWGN channel. Then, based on the achievable rate region of the GQF scheme, both the common and individual outage probabilities and the expected sum rate were derived. The numerical examples were presented to show the significant gain obtained by the GQF scheme over the classic CF scheme when the relay has no access to the CSI of the relay-destination link.

\section*{Appendix: Proof of Theorem \ref{th-QFD-MARC}}
Assume the source messages $W_{1}$ and $W_{2}$ are independent of each other. Each message $W_{i}$, $i\in\{1,2\}$, is uniformly distributed in its message set $\mathcal{W}_i = [1 : 2^{lR_i}]$.

\subsubsection{Codebook Generation}
Assume the joint pmf factors as
\begin{equation}
p(x_{11})p(x_{21})p(x_{12})p(x_{22})p(x_R)p(\hat{y}_R|y_R)
p(y_{D1},y_R|x_{11},x_{12})p(y_{D2}|x_{12},x_{22},x_R).
\end{equation}
Fix any input distribution $p(x_{11})p(x_{21})p(x_{12})p(x_{22})p(x_R)p(\hat{y}_R|y_R)$,
for $k=1,2$, randomly and independently generate
\begin{itemize}

    \item $2^{lR_k}$ codewords $x_{k1}^{n}(w_k)$, $w_k\in\mathcal{W}_k$, each according to $\prod _{i=1}^{n} p_{X_{k1}} (x_{k1,i}(w_k))$;
    \item $2^{lR_k}$ codewords $x_{k2}^{m}(w_k)$, $w_1\in\mathcal{W}_k$, each according to $\prod _{i=1}^{m} p_{X_{k2}} (x_{k2,i}(w_k))$;
    \item $2^{lR_U}$ codewords $x_{R}^{m}(u)$, $u\in\mathcal{U}=\{1,2,\dots 2^{lR_U}\}$, each according to $\prod _{i=1}^{m} p_{X_{R}} (x_{R,i}(u))$.
\end{itemize}
Calculate the marginal distribution
$$p(\hat{y}_R)=\sum_{x_{11}\in \mathcal{X_{11}} ,x_{21}\in \mathcal{X_{21}},y_{D1}\in \mathcal{Y_{21}},y_{R}\in \mathcal{Y_R}}p(\hat{y}_R|y_R)p(y_R,y_{D1}|x_{11},x_{21})p(x_{11})p(x_{21}),$$
randomly and independently generate $2^{lR_U}$ codewords $\hat{y}_{R}^{n}(u)$, each according to $\prod _{i=1}^{n} p_{\hat{Y}_{R}} (\hat{y}_{R,i}(u)).$

\subsubsection{Encoding}
To send message $w_i$, the source node $S_i$ transmits $x_{i1}^{n}(w_i)$ in the first slot and $x_{i2}^{m}(w_i)$ in the second slot, where $i\in\{1,2\}$. Let $\epsilon' \in (0,1)$ . After receiving $y_R^n$ at the end of the first slot, the relay tries to find a unique $u\in\mathcal{U}$ such that
\begin{equation}
(y_R^n,\hat{y}_R^n(u))\in \mathcal{T}_{\epsilon'}^n(Y_R,\hat{Y}_R)
\end{equation}
where $\mathcal{T}_{\epsilon}^n(Y_R,\hat{Y}_R)$ is the $\epsilon$-strongly typical set as defined in \cite{Lim2011}. If there are more than one such $u$, randomly choose one in $\mathcal{U}$. The relay then sends $x_R^m(u)$ in the second slot.

\subsubsection{Decoding}

The destination $D$ starts decoding the messages after the second slot transmission finishes. Let $\epsilon'<\epsilon<1$. Upon receiving in both slots, $D$ tries to find a unique pair of the messages $\hat{w}_1\in\mathcal{W}_1$ and $\hat{w}_2\in\mathcal{W}_2$ such that
\begin{eqnarray}
(x_{11}^n(\hat{w}_1),x_{21}^n(\hat{w}_2),y_{D1}^n,\hat{y}_R^n(u)) \in \mathcal{T}_\epsilon^n(X_{11},X_{21},Y_{D1},\hat{Y}_R)\nonumber\\
(x_{12}^m(\hat{w}_1),x_{22}^m(\hat{w}_2),x_R^m(u),y_{D2}^m) \in \mathcal{T}_\epsilon^m(X_{12},X_{22},X_R,Y_{D2})\nonumber
\end{eqnarray}
for some $u\in\mathcal{U}$.

\subsubsection{Probability of Error Analysis}
Let $W_i$ denote the message sent from source node $S_i, i \in \{1,2\}$. $U$ represents the message index chosen by the relay $R$.
Based on the symmetry of the codebook construction and the fact that the messages $W_i$  is chosen uniformly from $\mathcal{W}_i$, the probability of error averaged on $W_i$ and $U$ over all possible codebooks is
\begin{equation}
Pr(\mathcal\epsilon) = Pr(\hat{W}_1\neq 1 \cup \hat{W}_2\neq 1 | W_1=1, W_2=1)\label{poe}.
\end{equation}

Define two events $\mathcal{E}_{0}$ and $\mathcal{E}_{(w_1,w_2)}$:
{\setlength\arraycolsep{0.1em}
\begin{align}
\mathcal{E}_{0}  &:=  \{((Y_R^n,\hat{Y}_R^n(u))\notin  \mathcal{T}_{\epsilon'}^n(Y_R\hat{Y}_R)), \text {for all} \: u \}
\\
\mathcal{E}_{(w_1,w_2)}    &:=
\{ (X_{11}^n(w_1),X_{21}^n(w_2),Y_{D1}^n,\hat{Y}_R^n(u))
\in \mathcal{T}_{\epsilon}^n(X_{11}X_{21}Y_{D1}\hat{Y}_R) \:\: \text{and}
\nonumber \\
&\qquad (X_{12}^m(w_1),X_{22}^m(w_2),X_R^m(u),Y_{D2}^m)
\in \mathcal{T}_{\epsilon}^m(X_{11}X_{21}X_RY_{D2}) \; \text{for some}\: u \}.
\end{align}
}
Then $Pr(\mathcal\epsilon)$ can be rewritten as
\begin{eqnarray}
Pr(\mathcal\epsilon)
& \leq & Pr (\mathcal{E}_{0}|W_1=1,W_2=1)
+ Pr( (\mathcal{E}_{(1,1)} )^c\cap\mathcal{E}_{0}^c|W_1=1,W_2=1)
\nonumber \\
&  & + Pr(\cup_{(w_1,w_2)\in\mathcal{A}} \mathcal{E}_{(w_1,w_2)}|W_1=1,W_2=1),
\label{ineqn-err}
\end{eqnarray}

where $\mathcal{A}:=\{(w_1,w_2)\in\mathcal{W}_1\times \mathcal{W}_2:(w_1,w_2)\neq (1,1)\}$. Assume $\beta$ is fixed, then by covering lemma \cite{Gamal2010}, $Pr(\mathcal{E}_{0}|W_1=1,W_2=1)\rightarrow 0$ when $l\rightarrow \infty$, if
\begin{equation}
    R_U > \beta I(Y_R,\hat{Y}_R) + \delta(\epsilon')
\end{equation}
where $\delta(\epsilon')\rightarrow 0$ as $\epsilon'\rightarrow 0$. By the conditional typicality lemma \cite{Gamal2010}, $Pr( (\mathcal{E}_{(1,1)} )^c\cap\mathcal{E}_{0}^c|W_1=1,W_2=1) \rightarrow 0$ as $l\rightarrow \infty$. Through some standard probability error analysis \cite{Yao2013}, the second line of (\ref{ineqn-err}),$Pr(\cup_{(w_1,w_2)\in\mathcal{A}} \mathcal{E}_{(w_1,w_2)}|W_1=1,W_2=1)\rightarrow 0$, for fixed $\beta = \frac{n}{l}$,  $1-\beta = \frac{m}{l}$, if $l\rightarrow\infty$, $\epsilon\rightarrow 0$ and the inequalities (\ref{eqn-GQF-R1})-(\ref{eqn-GQF-R1R2RU}) hold. Therefore, the probability of error $P(\mathcal\epsilon) \rightarrow 0$. The proof completes and the achievable rate region is shown in \emph{Theorem \ref{th-QFD-MARC}}.

\bibliography{reference}
\bibliographystyle{IEEEtran}

\end{document}